# Bright high-purity quantum emitters in aluminium nitride integrated photonics

Tsung-Ju Lu[1,4*], Benjamin Lienhard[1,4], Kwang-Yong Jeong[1,3], Hyowon Moon[1], Ava Iranmanesh[1], Gabriele Grosso[2], and Dirk Englund[1*]

[1] Department of Electrical Engineering and Computer Science, Massachusetts Institute of Technology, Cambridge, MA 02139, USA
[2] Photonics Initiative, Advanced Science Research Center and Physics Program, Graduate Center, City University of New York, New York, New York 10031, United States
[3] Present address: Department of Physics, Korea University, Seoul 02841, Republic of Korea
[4] These authors contributed equally to this work
* E-mails: tsungjul@mit.edu; englund@mit.edu

**Abstract**

Solid-state quantum emitters (QEs) are fundamental in photonic-based quantum information processing. There is strong interest to develop high-quality QEs in III-nitride semiconductors because of their sophisticated manufacturing driven by large and growing applications in optoelectronics, high voltage power transistors, and microwave amplifiers. Here, we report the generation and direct integration of QEs in an aluminium nitride-based photonic integrated circuit platform. For individual waveguide-integrated QEs, we measure an off-chip count rate exceeding $6 \times 10^4$ counts per second (cps) (saturation rate $> 8.6 \times 10^4$ cps). In an unpatterned thin-film sample, we measure antibunching with $g^{(2)}(0) \sim 0.05$ and photon count rates exceeding $8 \times 10^5$ cps (saturation rate $> 1 \times 10^6$ cps). Although spin and detailed optical linewidth measurements are left for future work, these results already show the potential for high-quality QEs monolithically integrated in a wide range of III-nitride device technologies that would enable new quantum device opportunities and industrial scalability.

**Main Text**

Quantum emitters (QEs) with on-demand single photon emission are central building blocks for photonic-based quantum communication and quantum computation[1,2]. Over the past decade, crystal defect centres in wide-bandgap semiconductors have emerged as excellent solid-state QEs. In particular, a variety of colour centres in diamond[3–7], 4H silicon carbide (SiC)[8,9], and 6H SiC[10] have been demonstrated to have stable optical transitions coupled to long-lived spin ground states at room and cryogenic temperatures, resulting in spin-dependent single photon emission. The spin-photon interface provides access to electron spin states or indirectly to nuclear spin states, which can serve as quantum memories for quantum information processing[11,12] and quantum-enhanced sensing[13].

Although these QEs in diamond and SiC have the leading properties amongst solid-state emitters, there is a lack of active chip-integrated photonic components and wafer-scale thin film single-crystal diamond or SiC on low-index insulator, which limit the scalability of monolithic quantum information processing architectures in these materials. Hence, there is increased interest in exploring QEs in other material systems that can support both high-quality QEs and monolithic integration of wafer-scale photonic integrated circuits (PICs). Recently, alternative wide-bandgap materials such as two-dimensional hexagonal boron nitride (2D hBN)[14], gallium nitride (GaN)[15], and aluminium nitride (AlN)[16] have attracted attention as potential host materials for quantum emitters due to their single-crystal heteroepitaxy of thin film materials or the possibility to be integrated to any PIC platform in the case of 2D hBN[17]. Theoretical calculations show that AlN can serve as a stable environment for hosting well-isolated QEs with optically addressable spin states[18], though experimental demonstrations of such QEs are outstanding. In contrast to diamond or SiC, which have strong covalent bonds, AlN is an ionic crystal with piezoelectric properties that may offer strain-based acoustic control for quantum spins[19].

AlN is widely used in optoelectronics[20], high power electronics[21], and microelectromechanical systems[22], resulting in a large, continuously expanding industry bolstered by mature fabrication and growth technologies. AlN's optical transparency is bounded at the short-wavelength side by an exceptionally large bandgap of 6.015 eV (corresponding to ultraviolet) and extends to the mid-infrared spectrum[23]. Its high thermal conductivity ($\kappa$ = 285 W/m·K) and small thermo-optic coefficient (d$n$/dT = 2.32 × $10^{-5}$/K) enable AlN devices to operate with long-term optical and physical stability[24]. The wide bandgap and the availability of highly crystalline AlN thin films grown on low-index sapphire substrates, has enabled excellent AlN-based PIC platforms[20,25], with a wide variety of nonlinear optical effects such as parametric frequency conversion[26], sum/difference frequency generation[27], electro-optic modulation[24], and frequency comb generation[27]. Here, we report on two types of bright room-temperature QEs emitting in the orange visible spectrum integrated in a scalable AlN-on-sapphire PIC platform, as shown conceptually in Fig. 1a, made possible by overcoming several material and processing challenges. Well-established fabrication processes for AlN PICs shown in previous works[17,25,28] enable engineering of QE-based quantum photonic devices tailored to any application needs.

**Results**

**Material processing and quantum emitter creation in AlN**
Our studies used wurtzite-phase AlN thin film on sapphire ($Al_2O_3$) (Kyma Technologies, Inc.) with an AlN thickness of 200 nm. These samples were grown by plasma vapor deposition of nano-columns (PVDNC) with a macro defect density of less than 10 per $cm^2$. The crystal orientation is along the c-axis (00.1 ± 0.2°), as indicated in the black inset in Fig. 1a.

To create vacancy-based emitters, we ion-implanted the sample using a He ion microscope (Zeiss ORION NanoFab) at a He ion fluence of around $10^{15}$ ions/$cm^2$ with an accelerating voltage of 32 keV, followed by annealing in an argon environment at ambient pressure (see Methods), which prevents AlN surface oxidation that occurs in ambient air at temperatures above 700°C[29].

We investigated different annealing recipes with maximum temperatures between 500°C and 1000°C. As seen from the atomic force microscope scans in Fig. 1b, the 1000°C annealing reduced the root mean square surface roughness five-fold from 2.724 nm to 0.541 nm. The nano-columns maintain the hexagonal crystal structure after annealing, indicated in the cutouts in Fig. 1b, suggesting that the sample polytype remained wurtzite.

We characterized these samples by photoluminescence (PL) imaging in a confocal microscope setup, using 532 nm laser excitation (Coherent Verdi G5 laser, power before the objective ~1 mW) through a 0.9 NA microscope objective (Nikon). The PL was filtered with a 560 nm long-pass filter and detected with avalanche photodiodes (APDs by Excelitas) via single mode fibers. These measurements showed a spatially uniform PL background, but no resolvable isolated QEs, for AlN samples annealed up to 700°C; this background PL is reduced more than a hundred-fold and QEs are detectable in samples annealed at 1000°C, which is the material processing condition used for the samples that are studied in the rest of this work.

The observed PL background reduction in combination with the lower surface roughness suggests an improved crystal structure, characterized by higher-uniformity nano-columns. We tentatively attribute the reduced background to fewer unwanted defect centres in the bulk and on the surface of the material.

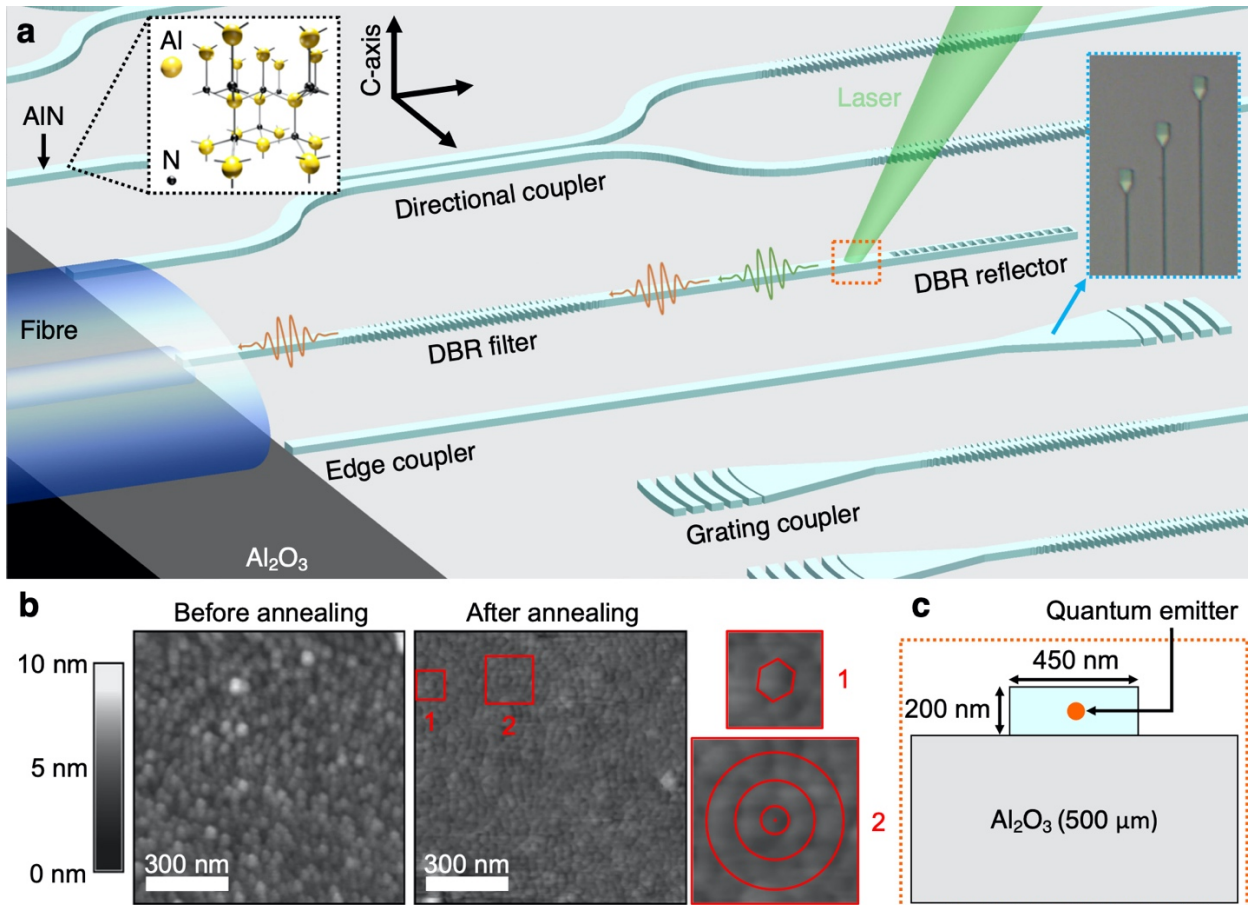

**Fig. 1 Quantum emitters in aluminium nitride integrated photonics. a** Scalable AlN-on-sapphire photonic integrated circuits with integrated quantum emitters. Black inset: Wurtzite crystal structure of aluminium nitride (yellow: aluminium atom, black: nitrogen atom). Blue inset: Microscope image of the fabricated QE-integrated waveguides, where the grating couplers are used for visual feedback during fibre edge coupling. **b** Atomic force microscopy of a sample before and after annealing. Cutout 1 indicates the hexagonal structure of the nano-columns is maintained after annealing. Cutout 2 shows slight coalescing of the AlN film columnar structure and improved orientation alignment to the c-axis, indicating an improved crystallinity to the AlN film. **c** Close-up cross-section of the single-mode AlN-on-sapphire waveguide, which is 450 nm in width by 200 nm in height. The quantum emitter is embedded within the AlN waveguide (not necessarily in the exact centre as shown).

**Spectral Analysis of Quantum Emitters**

The room-temperature PL scan of a representative AlN sample after 1000°C annealing (Fig. 2a) shows a dark background with bright, micron-spaced bright spots. Analysis of the PL spectra, photon statistics, emission lifetime, and polarization reveal these spots to be two orthogonally polarized classes of QEs labeled "A" and "B" (see Fig. 4d), as detailed next.

At room temperature, both types of QEs emit in a broad spectrum covering from 580 to 650 nm, as seen in the curves labeled as 295 K in Fig. 2b,d (spectra taken with Princeton Instruments Isoplane SCT320 spectrometer with a resolution of $0.07 \pm 0.01$ nm). The lines at 1.79 eV are caused by Chromium ion related impurities in sapphire[30] that we filter in following measurements with a 690 nm shortpass filter.

We performed low-temperature PL measurements to study phonon coupling characteristics in a closed-cycle helium cryostat (Montana Instruments) with a built-in confocal objective (NA = 0.9), and we used the same 532 nm laser for excitation and APDs for detection as previously stated. Fig. 2b,d shows low-temperature PL spectra of a type A and B QE, respectively, at 5 and 150 K. Common to all 5 K spectra is a strong zero-phonon line (ZPL) peak, accompanied by a red-shifted satellite peak.

The temperature dependence of the ZPL linewidth gives information about the interaction between the defects and host crystal lattice, as well as the emitter dephasing mechanisms. Fig. 2c shows the temperature dependence of the emitter linewidth approximated from Lorentzian fits for a type B emitter. We determine a linewidth of 0.16 ± 0.01 nm at 5 K, which is not limited by the spectrometer resolution (~0.07 nm). A $T^3$ function fits the linewidth broadening with increasing temperature, similar to the silicon vacancy (SiV) and other defects in diamond[31]. The $T^3$ dependence results from field fluctuations caused by phonon-induced dislocations of crystal defects and colour centres[32], allowing us to conclude that these AlN QEs are point defects.

The Debye-Waller factor (DWF) indicates the extent of electron-phonon coupling for the emitter. We estimate it from DWF = $I_{ZPL}/I_{TOT}$, where $I_{ZPL}$ is the ZPL PL emission and $I_{TOT} = I_{ZPL} + I_{PSB}$ is the total PL emission calculated by combining the ZPL PL emission $I_{ZPL}$ with the phonon broadened PL $I_{PSB}$. Here, we fit the ZPL and PSB peaks with separate Lorentzian fit functions. At cryogenic temperatures we determine the DWF to be 15 ± 2%.

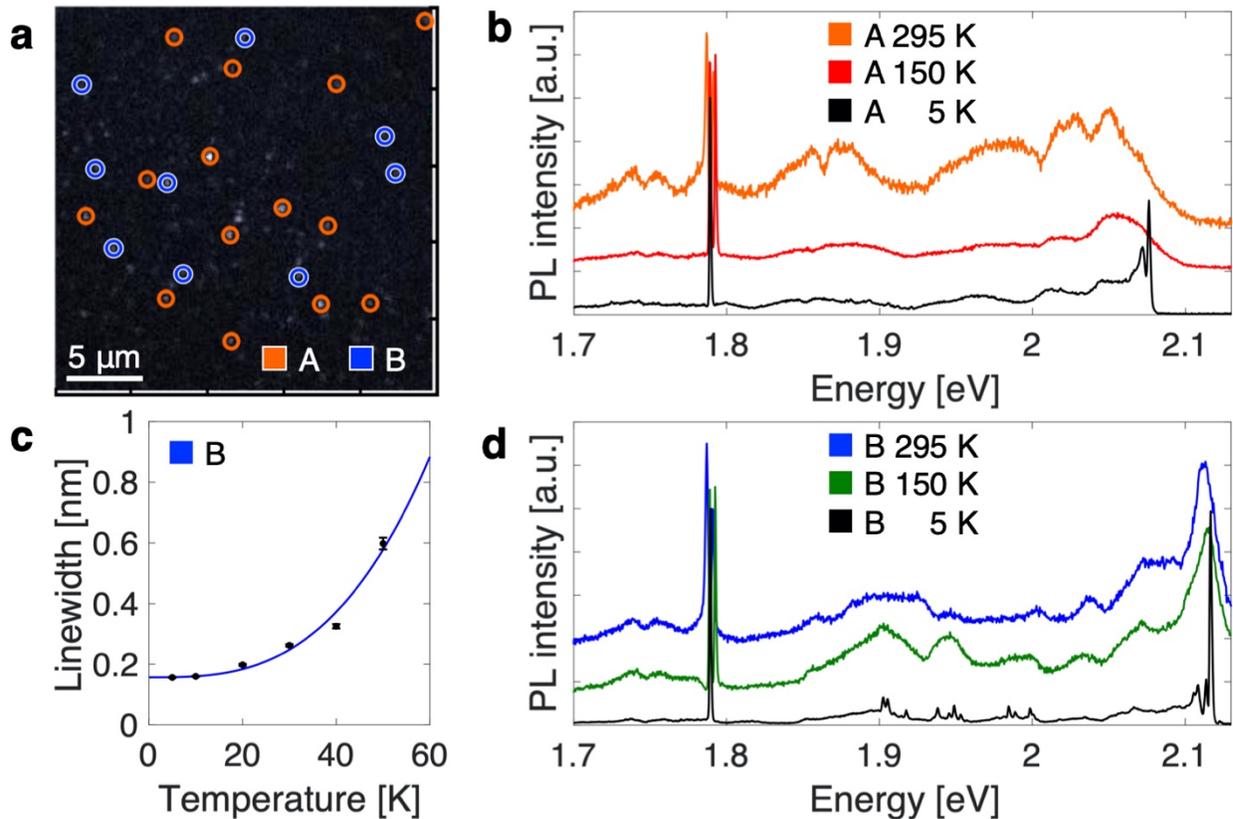

**Fig. 2 Spectral analysis of quantum emitters in thin-film w-AlN. a** Quantum emitter density in a 25 μm × 25 μm area (white scale bar: 5 μm). Two types of emitters with orthogonal polarization states are identified, labeled as "A" and "B". **b** Temperature-dependent PL spectra of a type A quantum emitter (black: 5 K, red: 150 K, orange: 295 K). **c** Temperature dependence of the type B quantum emitter's zero-phonon line linewidth. **d** Temperature-dependent PL spectra of a type B quantum emitter (black: 5 K, green: 150 K, blue: 295 K).

**Photon Statistics Characterization of Emitters**
We measured the second-order autocorrelation photon statistics of the emitters with a Hanbury Brown Twiss (HBT) interferometer. Fig. 3a,b show the normalized second-order autocorrelation histograms for representative type A and B emitters, respectively, under 532 nm, 1 mW excitation without background correction. The type A emitter has $g^{(2)}(0) = 0.08 \pm 0.06$ while the type B emitter has $g^{(2)}(0) = 0.09 \pm 0.08$, confirming that the emission is predominantly single photon emission for both types of emitters ($g^{(2)}(0) <$ 0.5). The histogram in Fig. 3a indicates a strong bunching near $\tau = 0$, whereas the histogram in Fig. 3b has weak bunching. This photon bunching feature suggests the presence of a dark metastable state. Fig. 3c shows second order autocorrelation histogram of a representative type A emitter under pulsed excitation with the laser set at a central frequency of 532 nm, 10 nm bandwidth, and 26 MHz repetition rate. Pulsed excitation resulted in further improved single photon purity of $g^{(2)}(0) = 0.05 \pm 0.002$.

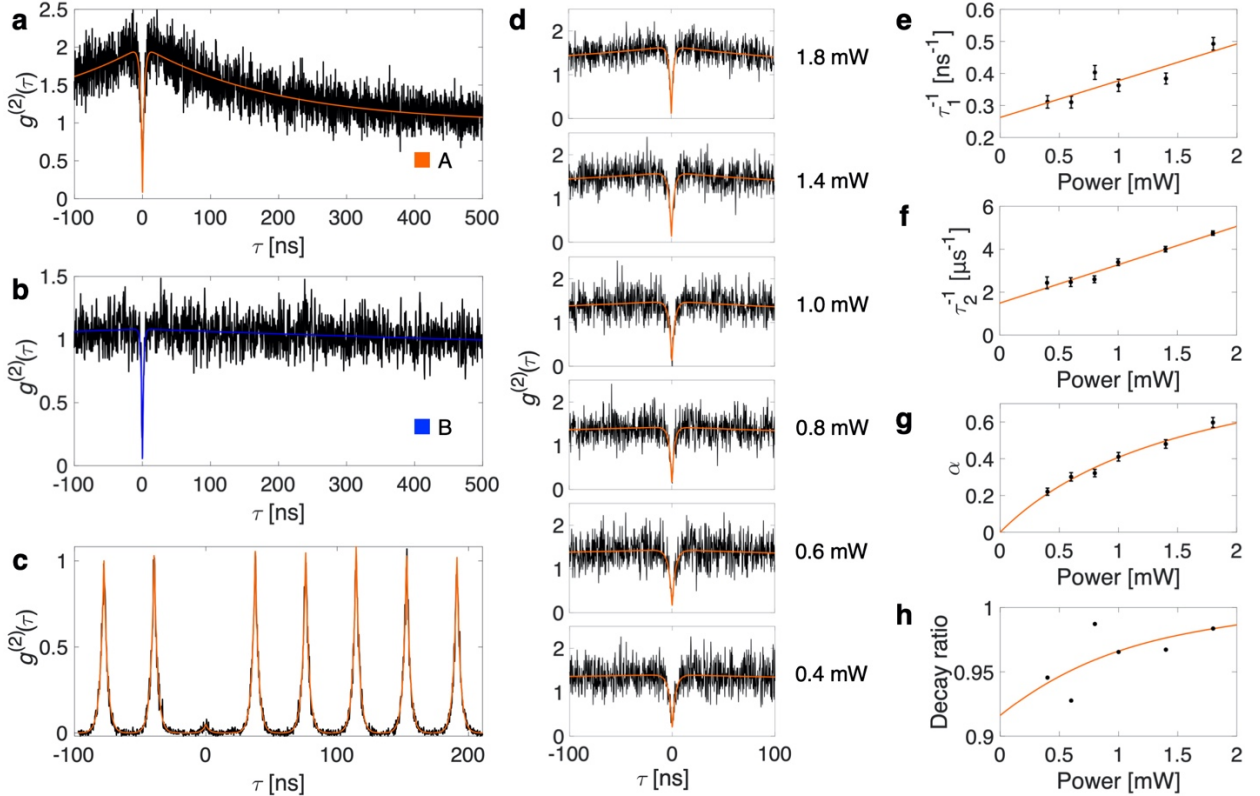

**Fig. 3 Photon statistics characterization of quantum emitters.** Continuous wave excitation second order autocorrelation histogram with **a** $g^{(2)}(0) = 0.08 \pm 0.06$ for a type A emitter and **b** $g^{(2)}(0) = 0.09 \pm 0.08$ for a type B emitter. **c** Pulsed excitation second order autocorrelation histogram for a type A emitter with $g^{(2)}(0) = 0.05 \pm 0.002$. **d** Power-dependent second-order autocorrelation histograms for a type A emitter. Power-dependence of the reciprocal values of fitting parameters **e** $\tau_1$ and **f** $\tau_2$ in the second-order autocorrelation fits. The plots are linearly fitted (orange) and reflect the decay rates of the excited and metastable states, respectively. **g** Power-dependence of the fitting parameter $\alpha$, which is representative of the non-radiative transitions via the metastable state. **h** Power-dependence of the decay ratio, which is defined as the ratio between the decay rate from the excited state to the ground state compared to the sum of all first order decay rates from the excited state.

A typical two-level model does not explain the bunching behavior near $\tau = 0$ in the second-order autocorrelation histograms. We therefore adopt the next-simplest level diagram shown in Fig. 4a: a three-level system with pump-power dependent transition rates. This models the essential features we observe and allows comparison with other well-studied emitters. Eq. (1) shows the three-level model for fitting the experimental $g^{(2)}(\tau)$ data:

$$g^{(2)}(\tau) = 1 - (1 + \alpha)\exp(-\frac{|\tau|}{\tau_1}) + \alpha\exp(-\frac{|\tau|}{\tau_2}) \qquad (1)$$

where $\tau_1$, $\tau_2$, and $\alpha$ are the excited state lifetime, metastable state lifetime, and fitting parameter. Fig. 3d shows the power-dependent second-order autocorrelation histograms for a type A emitter, which are used to evaluate the electron dynamics. We mainly focus on a type A emitter for these measurements due to its photostability compared to type B emitters, as shown in Fig. 4e.

Fitting parameter $\tau_1$ describes a two-level system with transitions between the ground state and a single excited state. Fig. 3e shows a plot of $1/\tau_1$ vs. excitation power, which can be linearly approximated as indicated by the orange fit. $\tau_1$ at zero excitation power gives the lifetime of the excited state[33,34]. Hence, the linear fit in Fig. 3e yields a lifetime of about 3.8 ns. The second fitting parameter $\tau_2$ describes the metastable state behavior. $\tau_2$ at zero excitation power gives the lifetime of the metastable state, which we found from the fit in Fig. 3f to be about 673 ns. The last parameter $\alpha$, plotted as a function of excitation power in Fig. 3g, is representative of the amount of bunching caused by the metastable state non-radiative transition. The stronger the bunching, the higher the probability of intersystem crossing. Furthermore, smaller the $\alpha$, the more the emitter behaves as a two-level system.

Fig. 3h shows as a function of excitation power the decay ratio, defined as the ratio of the decay rate from the excited state to the ground state versus all first order decay rates from the excited state (decay rate from the excited state to the ground state and decay rate from the excited state to the metastable state). The decay ratio from the excited state to the metastable state is constant for an ideal three-level model. Our experimental decay ratio versus power agrees with that, showing that the three-level system describes the type A emitter well and higher order levels transitions can be neglected.

**Photophysical Characterization of Emitters**
We use fluorescence lifetime measurements to compare with the lifetime values found previously to see how well the three-level system models our emitters. We can use the single exponential equation $I(t) \sim \exp(-t/\tau)$ for fitting the time-dependent PL intensity for the single state decay, in which $\tau$ represents the lifetime of the excited state. Fig. 4b shows the excited state lifetime measurement using a pulsed excitation laser with central wavelength of 532 nm, 10 nm bandwidth, pulse length of 1 ns, and 39 MHz repetition rate. The single-exponential fit indicates an excited state |e⟩ lifetime of 3.1 ± 0.1 ns for the type A emitter and 1.7 ± 0.1 ns for the type B emitter. The measured lifetime of the type A emitter is consistent with the lifetime values extrapolated from the power dependent $g^{(2)}(\tau)$ measurements, supporting that a three-level system is a suitable model for the type A emitters.

Fig. 4d shows the polar plots of the excitation-polarization-dependent PL intensity (without background subtraction) of both types of emitters fitted with a quadratic sinusoidal fit function $\sin^2(\theta + \varphi)$, where the angular parameter θ represents the rotation of the linearly polarized pump laser and $\varphi$ represents the orientation of the emitter relative to an arbitrary axis. Both types of emitters are shown to be single linearly polarized dipoles. Type A emitters have excitation and emission polarizations orthogonal to each other, while type B emitters have excitation and emission polarizations parallel to each other.

We next measured the PL intensity of the emitters as a function of the excitation power (with optimized polarization) to find the emitters' maximum emission rates. We used a second room temperature confocal microscope setup for these measurements, designed for efficient PL collection by using a Nikon 1.3 NA oil immersion objective for excitation and collection, as well as using a dichroic filter with 552 nm cut-off wavelength and a 560 nm longpass filter to remove the excitation laser and Raman lines. Fig. 4c (top) shows the saturation behavior of a type A emitter and the associated background.

Eq. 2 describes the typical function for fitting the emission count rate $I$ as a function of the measured excitation power $P$ for a three-level system:

$$I(P) = I_\infty \frac{P}{P_{sat} + P} + \alpha P + \beta \tag{2}$$

where $I_\infty$ is the emission rate at saturation, $P_{sat}$ is the excitation power at saturation, $\alpha$ is the linear power-dependent background slope, and $\beta$ is the constant dark counts. We use Eq. 2 to fit the intensity saturation behavior measurement plotted in Fig. 4c (top), which shows that the PL intensity exceeds $1 \times 10^6$ cps at saturation, with a saturation power of 1.5 mW. Fig. 4c (bottom) shows the $g^{(2)}(0)$ of the corresponding second-order autocorrelation measurements for each of the excitation powers used in Fig. 4c (top), indicating single photon emission characteristics for the emitter up to excitation powers of 3 mW, or twice the saturation power.

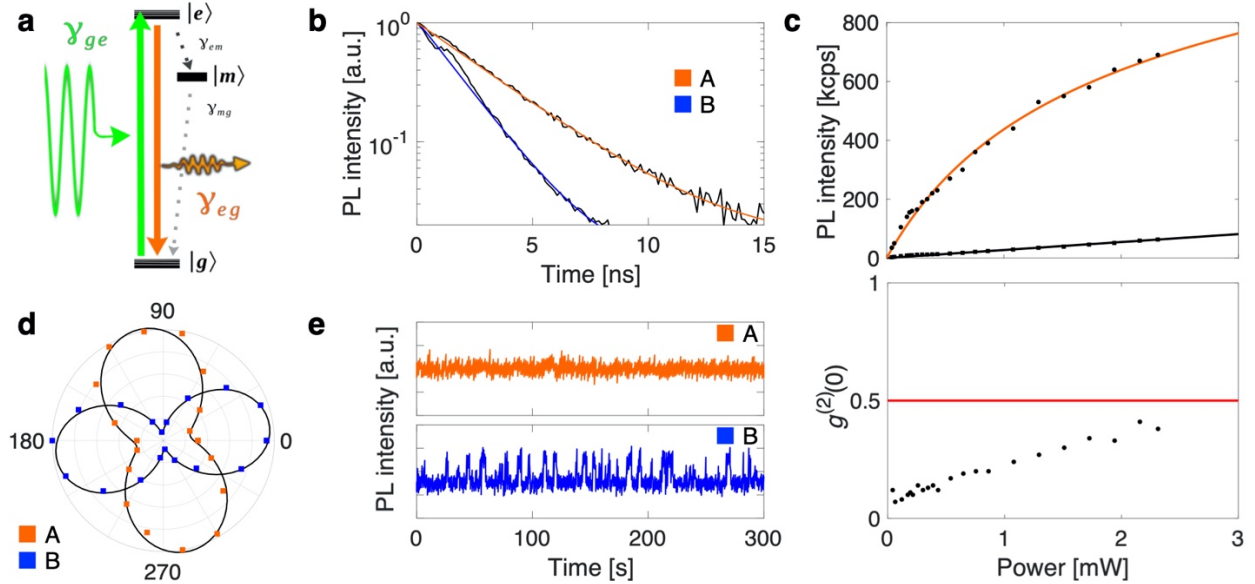

**Fig. 4 Photophysical properties of emitters. a** Graphical illustration of a three-level system. **b** Excited state lifetime measurements fitted with a single exponential decay function, showing an excited state lifetime of $3.1 \pm 0.1$ ns for type A emitters (orange) and $1.7 \pm 0.1$ ns for type B emitters (blue). **c** (top) PL intensity saturation response of a type A emitter exceeding 1 million counts per second at saturation, with a saturation power of 1.5 mW (data with orange fit). The data with black fit shows the associated background. (bottom) Plot of $g^{(2)}(0)$ as a function of excitation power, showing the high single photon emission purity up to twice the saturation power. Red line indicates the cutoff of $g^{(2)}(0) < 0.5$, indicating single photon emission. **d** Polar plots of PL as a function of linear excitation laser polarization. The emitters are split into two classes of emitters: one with a linearly oriented emission polarization orthogonal to the excitation spectrum (type A) and one with a linearly oriented emission polarization parallel to the excitation (type B). **e** Top: long-time photostability of a type A emitter. The emitter did not show any evidence of blinking or bleaching during the course of the experiments. Bottom: photostability of a type B emitter, showing blinking at sub-second timescales.

**Photonic Integration of Emitters**

Next, we created these QEs in AlN-on-sapphire PICs fabricated by a combination of electron beam lithography and dry etching (see Methods). The blue inset in Fig. 1a is a microscope image of the fabricated QE-integrated waveguides coupled off chip with edge couplers; the grating couplers are used for visual feedback during fibre edge coupling. Using the same room-temperature confocal microscopy setup with 0.9 NA objective described above for laser excitation, we used a cleaved Nufern UHNA3 fibre mounted on a XYZ piezo stage for edge-coupling to the AlN edge couplers, as illustrated conceptually in Fig. 1a. Our edge coupler design was not optimized to mode match with the UHNA3 fibre. We expect the experimental coupling efficiency to be less than 7% due to it being sensitive to the edge coupler polishing quality[28].

The confocal PL scan in Fig. 5a with confocal collection for detection of an AlN waveguide indicates QEs throughout, while Fig. 5b shows the same confocal PL scan with waveguide edge coupler collection into the UHNA3 fibre for detection. A comparison between the scans shows that all emitters are coupled into the waveguide. We use the circled emitter for the subsequent photon intensity correlation measurements.

We observe strongly antibunched emission from waveguide-coupled emission (Fig. 5d) and cross-correlation measurements (top versus waveguide collection, Fig. 5c). Despite the rather low estimated edge-coupling efficiency of 7%, we observe a high count rate exceeding $6 \times 10^4$ cps out of the waveguide (Fig. 5e), with $8.6 \times 10^4$ cps at a saturation power of 2.3 mW.

We show in Fig. 5f several exemplary devices realized on the same AlN-on-sapphire platform, though not characterized in this study. Distributed Bragg reflectors can be used as directional reflectors (red boxes) to direct all the emission towards one side of the photonic circuit, as well as spectral filters (green boxes) to filter out excitation light that is scattered into the waveguide. After the spectral filter, the emission can be either first split by an on-chip beamsplitter or directly coupled off the chip via edge couplers or grating couplers that can be placed anywhere throughout the chip (blue boxes).

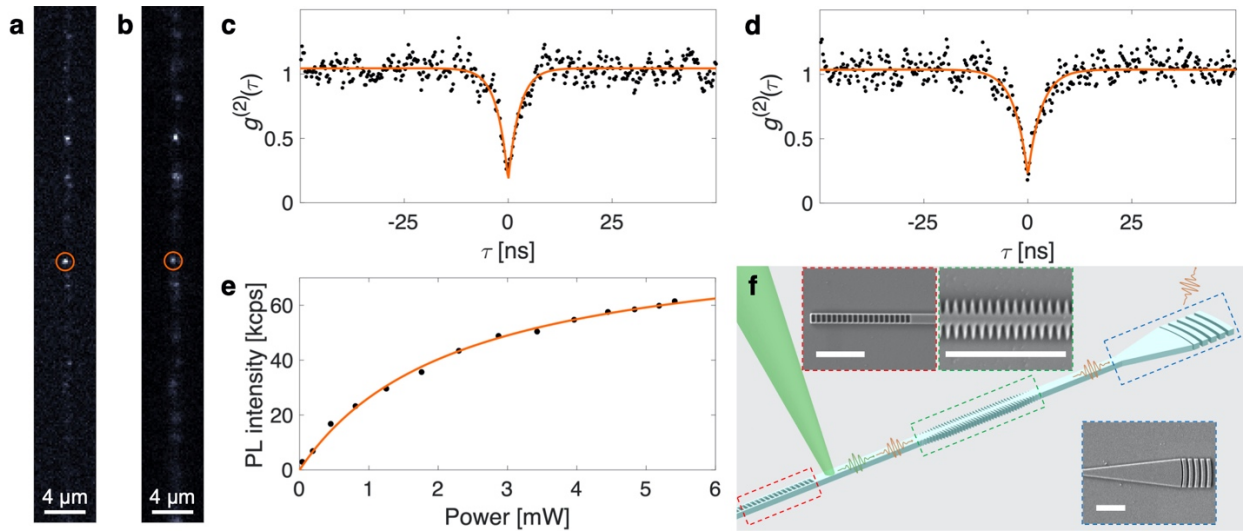

**Fig. 5 Photonic integration of emitters.** Confocal PL scan of AlN waveguide populated with quantum emitters throughout, with **a** confocal collection and **b** waveguide collection for detection. The circled emitter is the emitter that is studied in the photon-intensity correlation measurements. **c** Cross-correlation measurement of the emitter under 532 nm excitation between the photons collected via the confocal setup and the photons collected through the waveguide, with $g^{(2)}(0) = 0.17 \pm 0.07$, confirming the photons collected from the waveguide while exciting the emitter originates from the emitter. **d** Autocorrelation measurement of the emitter via waveguide collection only, with $g^{(2)}(0) = 0.21 \pm 0.08$. **e** PL intensity saturation response of an emitter with waveguide collection, with counts exceeding $8.6 \times 10^4$ cps at a saturation power of 2.3 mW. **f** Conceptual diagram of an AlN PIC with distributed Bragg reflectors as filter (green) and directional reflector (red), as well as grating coupler (blue) for dense population of read-out channels on the chip. Insets are SEM micrographs of the fabricated structures, with scale bars in each being 2 μm.

**Discussion**

While we observed the formation of the QEs after He-ion implantation and annealing, both type A and B QEs are also observed for samples that did not undergo He-ion implantation, albeit at a lower density, making it likely that the emitter type is either intrinsic or vacancy-based. Some sort of marker indicators on the bulk sample that are inert and do not degrade with the material treatments are needed to do a more careful study on the He-ion dose's effect on the QE formation and properties. As our emitters are in a 200 nm thick AlN film, we speculate from their stronger blinking behavior that type B emitters may be closer to the surfaces[14].

There have been many previous works reporting on quantum emitters in other nitride materials with emission wavelengths covering a wide wavelength range similar to the emitters in this work[35–37]. However, even with extensive theoretical density functional theory (DFT) modeling, the origin of these other emitter defects is not fully understood[38]. Our QEs in thin film AlN offer insights and potential future research directions to reveal the origins of similar QEs that have been observed in both monolayer and microns thick bulk nitride materials, such as the possibility of direct 3D atomic reconstruction by scanning transmission electron microscopy[39].

In conclusion, we demonstrated quantum emitters in a 200 nm thin-film AlN grown on top of sapphire, with count rates exceeding 1 million cps at saturation and high purity even at room temperature. We also showed that these QEs can be readily integrated in AlN-on-sapphire PICs using conventional AlN fabrication processes for patterning photonic components. While further studies are needed to investigate the predicted spin-dependent transitions coupling to spin qubits, strain-driven spin control, and other avenues towards spectrally narrower emission, the integration of stable QEs into AlN-on-sapphire PICs already opens the prospect of stable single photon sources integrated monolithically with optical modulators[24], AlN-integrated single photon detectors[40], and frequency conversion devices[41]. Furthermore, this work shows the potential of integrating high-quality QEs to a wide range of technologies comprised of the aluminium gallium nitride (AlGaN) family of materials that, thanks to their exceptional optical and electronic properties, as well as a large and growing industrial base, already form the state-of-the-art for applications in UV lasers, piezoelectric actuators and filters, high-power and high-speed electronics, and solid state lighting[42–44].

**Acknowledgements**
This work was supported in part by the Army Research Office MURI (Ab-Initio Solid-State Quantum Materials) grant number W911NF-18-1-0431 and in part by National Science Foundation (NSF) Research Advanced by Interdisciplinary Science and Engineering (RAISE), grant no. CHE-1839155. T.-J.L. acknowledges support from the Department of Defense National Defense Science and Engineering Graduate Fellowship as well as the Air Force Research Laboratory RITA program FA8750-16-2-0141. B.L. acknowledges support from Center for Integrated Quantum Materials (CIQM). K.-Y.J. acknowledges support from the National Research Foundation of Korea (NRF) grant funded by the Korean government (MSIT) (2015R1A6A3A03020926 and 2018R1D1A1B07043390). H.M. acknowledges support from the Samsung Scholarship as well as partial support from the NSF RAISE TAQS program. A.I. acknowledges support by the MIT UROP program. G.G. acknowledges support from the Graduate Center of the City University of New York (CUNY) through start-up funding.